\documentclass[showpacs,preprint,amsmath,amssymb]{revtex4}
\usepackage{graphicx,graphicx,epsfig}

\usepackage{graphicx}
\usepackage{bm}

\begin{document}
\title{Spin polarization control by electric stirring: proposal for a spintronic device}

\author{Yu. V. Pershin$^{1}\footnote{ equal contribution}$, N. A. Sinitsyn$^{2*}$, A. Kogan$^3$, A. Saxena$^4$, and D. L. Smith$^4$}
\affiliation{$^1$Department of Physics and Astronomy and USC
Nanocenter, University of South Carolina, Columbia, SC, 29208 USA
\\ $^2$Center for Nonlinear Studies and Computer, Computational
  and Statistical Sciences Division, Los Alamos National Laboratory,
  Los Alamos, NM 87545 USA\\ $^3$Department of Physics,
  University of Cincinnati, Cincinnati, OH USA \\
  $^4$Theoretical Division, Los Alamos National Laboratory,
  Los Alamos, NM 87545 USA}


\begin{abstract}
We propose a spintronic device to generate spin polarization in a
mesoscopic region by purely electric means. We show that the spin
Hall effect in combination with the stirring effect are sufficient
to induce measurable spin polarization in a closed geometry. Our
device structure  does not require the application of magnetic
fields, external radiation or ferromagnetic leads, and can be
implemented in standard semiconducting materials.
\end{abstract}

\date{\today}

\maketitle

Efficient control over electronic spins has a potential to
transform the design principles in semiconductor electronics.
Spins provide additional degrees of freedom, which can be used as
mobile bits of information and manipulated on much faster
time-scales than charges. Spintronic devices, such as the
giant-magnetoresistive effect based devices, have already found
industrial applications \cite{dyakonov-book}. However, realizing
the full potential of spintronics requires  integrating spin-based
devices with traditional semiconductor technology.

Various techniques have been proposed to manipulate spins in
semiconductors. The most accessible ones include the application
of an external magnetic field or polarized radiation
\cite{dyakonov-book}, but they do not have the potential for
implementing a single-chip device. Another possibility is to
generate spin-polarized currents by coupling semiconducting and
ferromagnetic materials. This approach, while promising, currently
faces  many technological complications \cite{dyakonov-book}.
Thus, it is most desirable  to create spintronic components which
involve neither external optical/magnetic fields nor coupling to
ferromagnetic materials for their operation. It is preferred that
such components be implemented in standard non-magnetic
semiconductors, such as Si or GaAs. For this reason, the idea of
using microwave fields in spintronic applications has attracted
much attention \cite{dyakonov-book, aronov, loss-nature,martin}.
However, these effects still rely on the presence of  external
magnetic fields to break the time-reversal symmetry or appear as
bulk effects for which applications to building new spintronic
components remain obscure.

In this Letter, we propose a device structure to generate spin
polarization in a mesoscopic region. In our setup, the
time-reversal symmetry is broken by purely electric means when
multiple  AC gate voltages, working with the phase shifts
different from $0$ or $\pi$, are applied. Periodic variation of
parameters in a closed conducting system leads to the stirring
effect (SE) \cite{stirring-quantum,sinitsyn-09review}, which is a
sort
 of pump effect \cite{niu-pump,vavilov-review}
with a distinctive feature that currents during SE are induced in
a closed geometry, without sinks or sources of electrons. These
currents have to be circulating.

\begin{figure} %[b]%{2.2in}%[t]
%\centerline{\includegraphics[width=4.5cm]{fig1}}
\centerline{\includegraphics[width=8cm]{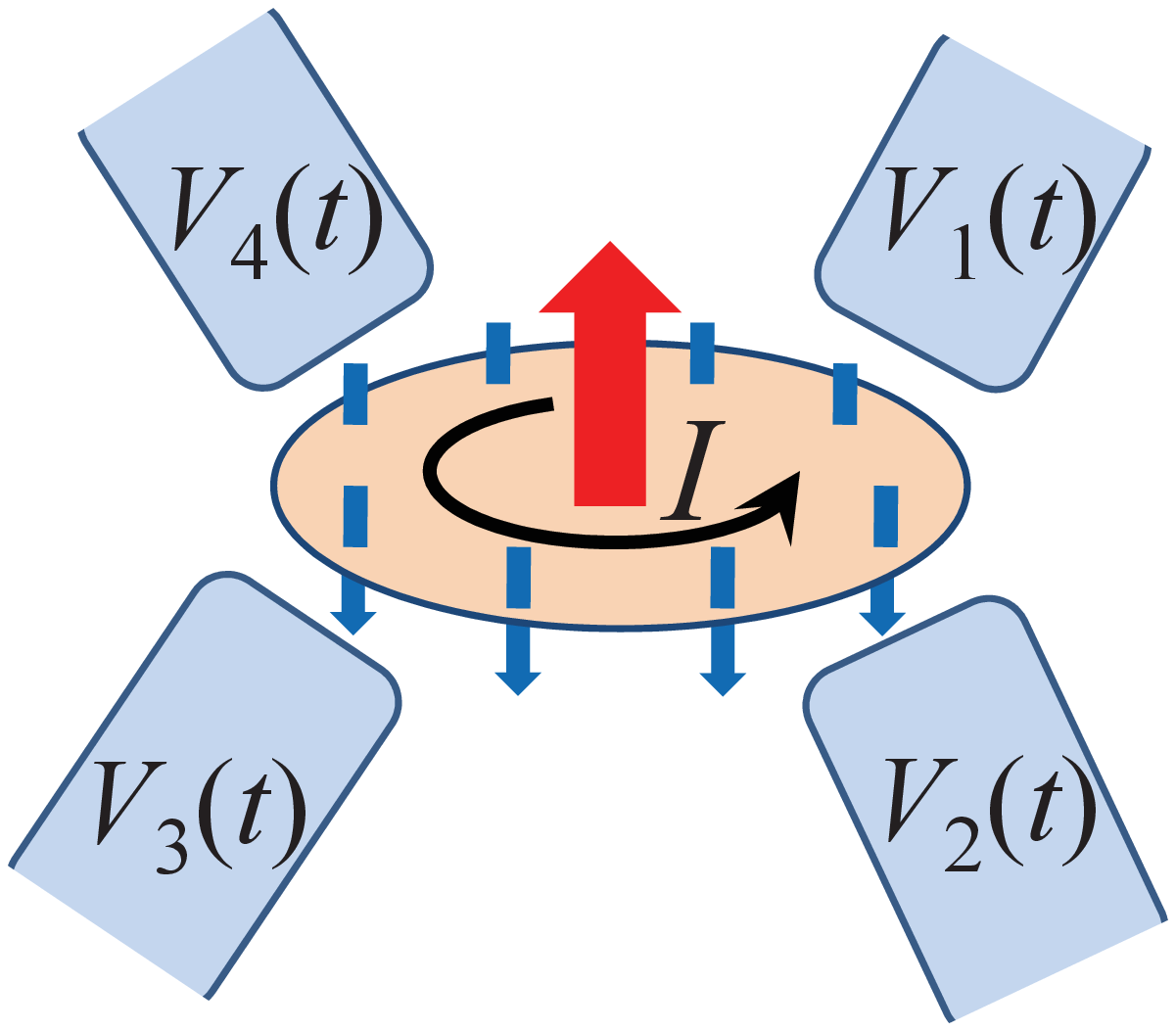}}
 \caption{(Color
online) Spintronic device to generate spin polarization at the
center of a semiconductor island. Application of AC gate voltages
induces circular charge currents in the island. Spin Hall effect
in such a system results in opposite spin accumulation at the
center and near the boundaries of the island. Direction of the
accumulated spins is along $z$ axis and determined by the
direction of gate-controlled circular charge currents.
\label{fig1}}
\end{figure}

Consider, for example, a mesoscopic conducting sample, which we
will call a {\it conducting island}. In Fig.~\ref{fig1}, four
gates with a voltage signal  changing with time according to the
law $$ V_k(t) = V \sin (\omega t+2\pi k/4),\quad k=1,..,4, $$ will
induce the SE so that the electric current will flow in a
preferred (clockwise or counterclockwise) direction inside the
island. Let the conducting island be made of a material exhibiting
the spin Hall effect
\cite{murakami-she,sinova-she,kato,wunderlich} (SHE). Then, when
circulating charge currents are excited, electrons with a certain
spin polarization will deflect towards the island's center as it
is shown in Fig. \ref{fig1} creating the desired spin
polarization.

We take the size of conducting island to be sufficiently large to
disregard the discreteness of electron energy spectrum. Depending
on the screening strength, the gate voltages can create
circulating currents in the whole island or only in a localized
region near the boundary. The former regime is more desirable and
is realizable in low or moderately doped semiconductor islands
whose dimensions do not exceed several $\mu$m (see calculations
below). The second regime is more relevant to heavily doped
semiconductor
or metal islands. % Due to the finite spin life time, the size of
%the setup cannot be arbitrarily large.
By solving drift diffusion equations near the interface between
regions with zero and nonzero charge currents, one can find that
spin polarization decays at a distance comparable to the spin
diffusion length $L_s=\sqrt{D\tau_s}$ from the interface, where
$D$ is the charge diffusion coefficient and $\tau_s$ is the spin
relaxation time.  To find substantial spin polarization near the
center of the island, in the second regime, the size of the
region, unperturbed by gate voltages, should not be much larger
than $L_s$. Typically, $L_s\sim 1-10$ microns for GaAs. Hence, due
to the finite spin life time, the size of the conducting island
can be several microns. In our numerical simulations (see below),
we consider a situation when the diameter of the island is smaller
than $L_s$. In this case, substantial spin polarization at the
island center is obtained.

We note that there exists another mechanism for spin polarization
in our setup. Microscopically, spin-orbit coupling is described by
the expression
$
E_{so} =g_{so}{\bf \hat{L} \cdot \hat{S}},
$
where $\hat{\bf L}$ is the operator of angular momentum and
$\hat{\bf S}$ is the spin operator and $g_{so}$ is a spin-orbit
coupling constant. Circulating currents carry a nonzero angular
momentum, and hence lead to energy imbalance for spins with
opposite orientations. This
 mechanism can be functional even in devices with
nanoscale dimensions. However, it is expected that the SHE is the
dominant mechanism for non-zero spin polarization, and, therefore,
only the SHE is included in our calculations.

In the drift-diffusion approximation, the z-component of spin
current ${\bf J}_s^z$ is given by
\begin{equation}
{\bf J}_s^z=-D{\bf \nabla}P_z+\lambda_{sh}\hat{z} \times {\bf
J}/e, \label{spin-current}
\end{equation}
where $P_z=n_\uparrow-n_\downarrow$ is the spin density imbalance,
$n_{\uparrow(\downarrow)}$ is the density of spin-up (-down)
electrons, $\hat{z}$ is a unit vector in the direction which is
transverse to the stirring plane and ${\bf J}$ is the charge
current density. The first term in the RHS is due to spin
diffusion, and the second term describes the SHE.
 Experimental\cite{kato} and theoretical\cite{engel} studies in GaAs
as well as Al samples determined the relative strength of the spin
Hall current to the charge current as $\lambda_{sh} \sim 10^{-3}-
10^{-4}$. We use a linear dependence of spin current on the charge
current, which describes the extrinsic mechanism of the SHE. It is
believed to dominate in n-doped GaAs \cite{kato}.

Consider a case of a  uniformly circulating electron current ${\bf
J}(r,\theta) =e\omega n r \hat{\theta}$,
 where $n$ is the density of conducting electrons,
 $\omega$ is the rotation frequency,
 and $\hat{\theta}$ is the unit vector in the azimuthal direction.
In the polar coordinates, the drift-diffusion equation with spin
currents (\ref{spin-current}) and a phenomenological spin
relaxation term is given by
\begin{equation}
\frac{\partial P_z}{\partial t}=D \left(\frac{\partial^2
P_z}{\partial r^2}+\frac{1}{r} \left( \frac{\partial P_z}{\partial
r} \right)\right) + \lambda_{sh}\frac{1}{r}
\frac{\partial}{\partial r} (rJ_{\theta})-\frac{P_z}{\tau_s},
\label{polar-diff}
\end{equation}
where $r$ and $\theta$ are coordinates of a point in the polar
coordinate system.
%In Eq. (\ref{polar-diff}) all terms are averaged over one period of the gate voltage oscillation.
  Eq. (\ref{polar-diff}) has a simple stationary solution that describes a rotating electron gas in a
conductor of infinite diameter and
\begin{equation}
P_z=2\omega n \lambda_{sh}\tau_s. \label{sol-1}
\end{equation}
It cannot be valid near the boundary of a finite system, however,
it provides a good estimate of the spin polarization {\it far away
from the boundary}  inside a sufficiently large structure. Taking
$\tau_s\sim 10^{-8}s$, $\lambda_{sh}= 10^{-4}$, and $\omega \sim
10^{9}s^{-1}$ we find that $P_z \sim n 10^{-3}$, i.e. about 0.1\%
of electrons will be spin polarized near the center of the
structure.

The application of an AC gate voltage is akin to the application
of a rotating electric field. At optical frequencies, such a field
is known to induce transitions between the valence and conduction
bands of semiconductors, creating spin polarization. In our
device, this field rotates with a microwave frequency and cannot
induce spin polarization via the conventional mechanism. It can,
however,
excite circulating currents %as it happens, e.g., in quantum rings
\cite{pershin-piermarocchi}.

We solve numerically a system of two-component drift-diffusion
equations supplemented by the Poisson equation. Our numerical
scheme is similar to those used in Ref.
\onlinecite{pershin-diventra} with the only difference that now we
are solving equations in 2D. The Poisson equation is solved in a
larger area enclosing the conducting island with the boundary
condition that the potential at the boundary of the larger area is
presented by the rotating electric field. The rotating electric
field can be written as $\boldsymbol{E}=E_0\cos (\omega
t)\hat{x}\pm E_0\sin (\omega t)\hat{y}$, where $E_0$ and $\omega$
are the electric field amplitude and angular frequency, $\hat{x}$
and $\hat{y}$ are unit vectors in the $x$ and $y$ directions, and
$\pm$ corresponds to a $\sigma_\pm$ circular polarization.

\begin{figure}%{2.2in}%[t]
\centerline{\includegraphics[width=8.0cm]{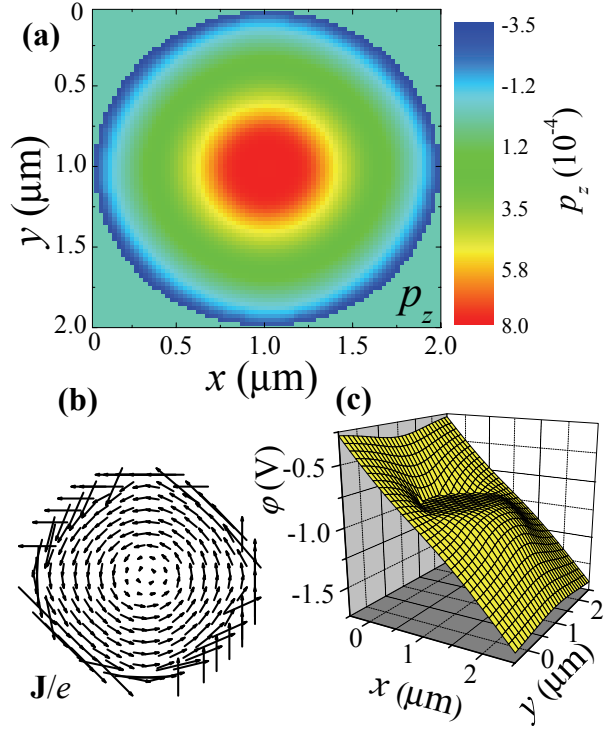}}
 \caption{(Color
online) Distributions of  (a) spin polarization $p_z$,  (b) charge
current density ${\bf J}/e$ and (c) potential calculated in a
system containing a circular semiconductor island of radius
$R=1\mu$m subjected to a rotating electric field. (a) and (b) were
found as an average over a field rotation period $T$, while (c) is
an instantaneous potential profile at $t=5$ns. The following
values of parameters were used: $n=10^{15}$cm$^{-3}$,
$\tau_{s}=20$ns, $\lambda_{sh}=10^{-3}$,
$\mu=8500$cm$^2$/(V$\cdot$s), $L_s=7\mu$m, $E_0=5$kV/cm, $T=1$ns.
$E_0=5$kV/cm corresponds to 1V voltage drop over 2$\mu$m (diameter
of the island).  \label{fig2}}
\end{figure}

Fig. \ref{fig2} shows a representative result of our calculations.
Here, we consider a moderately doped GaAs island of a circular
shape. At the selected value of electron density
($n=10^{15}$cm$^{-3}$), the electric field penetrates deep inside
the island creating circulating currents which are not limited to
the surface (Fig. \ref{fig2}(b)). The spin polarization calculated
as $p_z=\langle n_\uparrow-n_\downarrow \rangle / \langle
n_\uparrow+n_\downarrow \rangle$, where $\langle .. \rangle$
denotes averaging over the rotating field period, has a magnitude
($\sim 8\times 10^{-4}$) at the center.

\begin{figure}%{2.2in}%[t]
\centerline{\includegraphics[width=8cm]{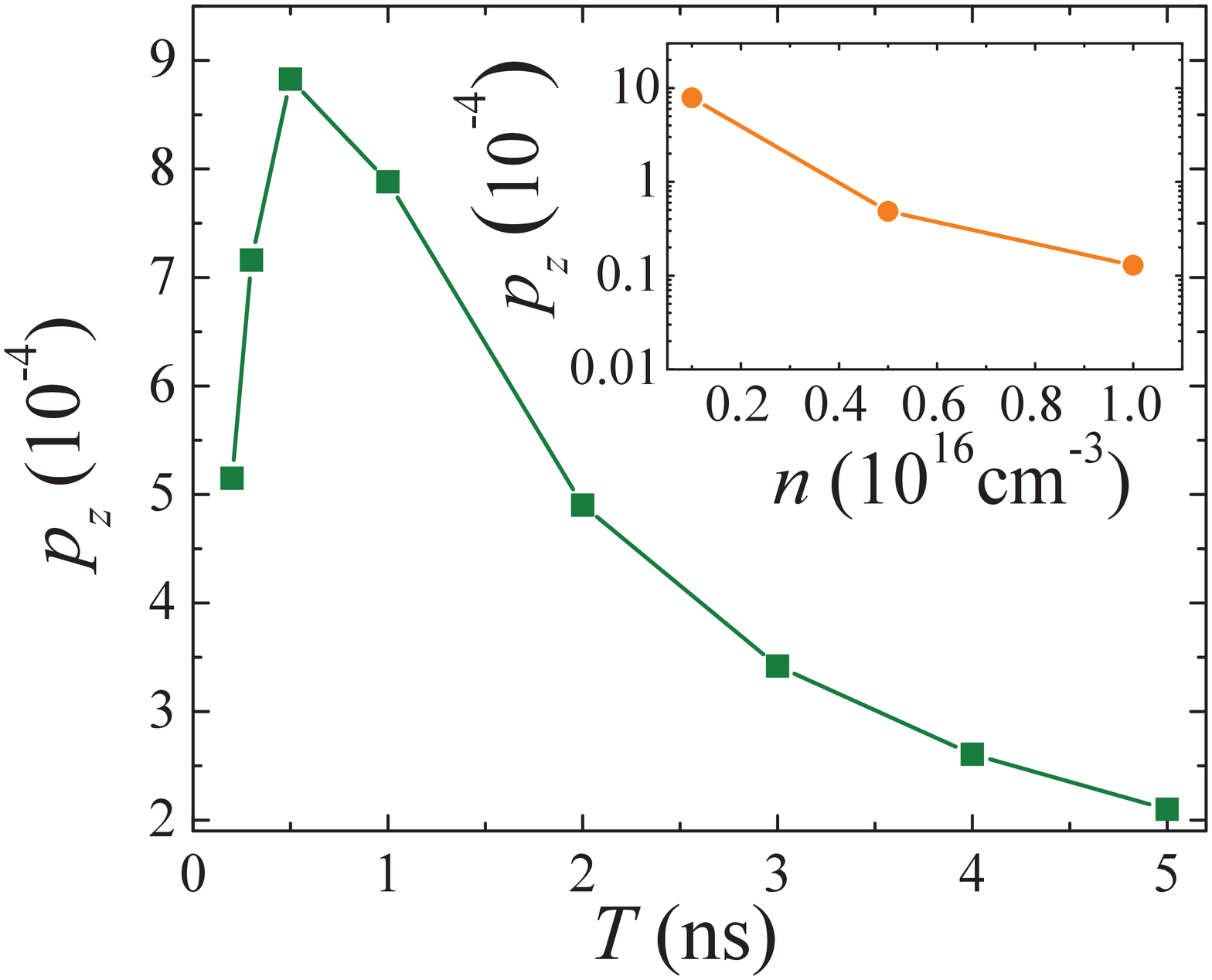}}
 \caption{(Color
online) Spin polarization at the island's center at different
values of the field rotation period $T$. Inset: spin polarization
at the island's center at different values of electron density $n$
at $T=1$ns. Here, we used the same values of all other parameters
as in Fig. \ref{fig2}. \label{fig3}}
\end{figure}

In Fig. \ref{fig3}, we plot the spin polarization magnitude at the
island center as a function of the rotating field period and
electron density (inset). When we start to decrease $T$ from 5ns,
$P$ increases first as is expected from Eq. (\ref{sol-1}), and has
a maximum at $T\sim 0.5$ns and then decreases. This decrease is
possibly related to the resonance nature of circulating current
excitation or strong spin densities mixing in highly
non-equilibrium environment.
 An increase of the electron density results in a
decrease of the spin polarization (see the inset in Fig.
\ref{fig3}) mainly because  at higher electron densities, the
circulating currents are limited to the surface regions. We note
that in a real setup, this effect should be less important because
the finite viscosity of the electron fluid as well as the finite
thickness of the island are not captured by our 2D calculations.
Then, when a finite thickness of the island is taken into account,
the exponential 3D screening reduces to a power-law type
\cite{stern} and the electric field penetrates deeper into the
island creating stronger SE.

In conclusion, we proposed a spintronic device that generates a
spin polarization in a localized region of a semiconductor by
purely electric means. Importantly, our device structure does not
require the application of magnetic fields, external radiation or
ferromagnetic leads, and can be implemented in standard
semiconductor materials. The  novelty  of our approach is in
considering the combination of two effects, namely the spin Hall
effect and the stirring effect. Our numerical simulations confirm
that the induced spin polarization is sufficiently strong to be
observable by the Kerr rotation technique.
% New spintronic effects can be studied upon
%realizing our spintronic device.
Our spintronic device can be used to generate spin polarized
currents or to control the magnetization of a nanomagnet placed at
the center of the conducting island, e.g. by doping Mn ions.

{\it Acknowledgments.} This work was funded in part by NSF grant
DMR 0804199, and the US DOE under Contract No. DE-AC52-06NA25396.

\newpage

%%%%%%%%%%%%%%%%%%%%%%%%%%%%%%%%%%%%%%%%%%%%%%%%%%%%%%%%%%%%%%%%%%%%%%%%%%%%%%%%%%%%%%%%%%%
%%%%%%%%%%%%%%%%%%%%%%%%%%%%%%%%%%%%%%%%%%%%%%%%%%%%%%%%%%%%%%%%%%%%%%%%%%%%%%%%%%%%%%%%%%%

%%%%%%%%%%%%%%%%%%%%%%%%%%%%%%%%%%%%%%%%%%%%%%%%%%%%%%%%%%%%%%%%%%%%%%%%%%%%%%%%%%%%%%%%%%%
%%%%%%%%%%%%%%%%%%%%%%%%%%%%%%%%%%%%%%%%%%%%%%%%%%%%%%%%%%%%%%%%%%%%%%%%%%%%%%%%%%%%%%%%%%%

%%%%%%%%%%%%%%%%%%%%%%%%%%%%%%%%%%%%%%%%%%%%%%%%%%%%%%%%%%%%%%%%%%%%%%%%%%%%%%%%%%%%%%%%%%%
%%%%%%%%%%%%%%%%%%%%%%%%%%%%%%%%%%%%%%%%%%%%%%%%%%%%%%%%%%%%%%%%%%%%%%%%%%%%%%%%%%%%%%%%%%%

\end{document}